\input amssym.def
\input amssym.tex
\def\ip{interstellar medium }
\def\gr{$\gamma$-ray }
\def\grs{$\gamma$-rays }
\def\be{\begin{equation}}
\def\ee{\end{equation}}
\def\bdm{\begin{displaymath} }
\def\edm{\end{displaymath} }
\documentclass{aa}
\usepackage{times}
\usepackage{graphics}

\begin{document}

\title{Excess GeV radiation and cosmic ray origin}
\author{I. B\"usching$^1$, M. Pohl$^2$, R. Schlickeiser$^3$}
\date{Received ? / Accepted ?}
\institute{Institut f\"ur Theoretische Physik, Lehrstuhl IV: 
Weltraum- und Astrophysik, Ruhr-Uni\-ver\-si\-t\"at Bochum, 
D-44780 Bochum, Germany; \\
$^1$ib@tp4.ruhr-uni-bochum.de; 
$^2$mkp@tp4.ruhr-uni-bochum.de; 
$^3$rsch@tp4.ruhr-uni-bochum.de}

\offprints{I. B\"usching}
%\date{ }
\authorrunning{B\"usching, Pohl, Schlickeiser}
\titlerunning{Excess GeV radiation}
\abstract{Particle acceleration at supernova remnant (SNR) shock waves is regarded as
the most probable mechanism for providing Galactic cosmic rays at energies
below $10^{15}$ eV. The Galactic cosmic ray hadron component would
in this picture result from the injection of relativistic
particles from many SNRs. It is well known
that the superposition of individual power law source spectra with dispersion
in the spectral index value, which behaviour is observed in the
synchrotron radio spectra of shell SNR, displays a positive curvature 
in the total 
spectrum and in particular shows a hardening at higher energies. 
Recent observations
made with the EGRET instrument on the Compton Gamma-Ray Observatory
of the diffuse Galactic $\gamma$-ray emission reveal a spectrum which is 
incompatible with the assumption that the cosmic ray spectra measured
locally hold throughout the Galaxy: the spectrum above 1 GeV, 
where the emission is supposedly dominated by $\pi^0$-decay, is harder
than that derived from the local cosmic ray proton spectrum.
We demonstrate that in case of a SNR origin
of cosmic ray nucleons part of this $\gamma$-ray excess may be attributed to 
the 
dispersion of the spectral indices in these objects. In global averages, as 
are $\gamma$-ray line-of-sight integrals, this dispersion leads to a positive
curvature in the composite spectrum, and hence to modified $\pi^0$-decay
$\gamma$-ray spectra.
}
\maketitle

\section{Introduction}
Particle acceleration at supernova remnant shock waves is regarded as
the most probable mechanism for providing Galactic cosmic rays at energies
below $10^{15}$ eV (for a review see Blandford \& Eichler \cite{be87}).
The recent detections of non-thermal X-ray synchrotron radiation
from the four supernova remnants SN1006 (Koyama et al.\cite{koyama95}), RX 
J1713.7-3946 (Koyama et al. \cite{koyama97}), 
Slane et al. \cite{sla99}), Cas A (Allen et al. 
\cite{allen97}), and RCW86 (Borkowski et al. \cite{bor01}) and the 
subsequent detections of SN1006 (Tanimori et al. \cite{ta98})
and RX J1713.7-3946 (Muraishi et al. \cite{mur00}) at TeV energies 
support the hypothesis that at least Galactic cosmic ray electrons are 
accelerated predominantly in SNR. To date, there is still no unambiguous
proof that cosmic ray nucleons are similarly produced in SNR.

Whatever the nature of their sources, upon escape from there the cosmic ray 
nucleons would
diffusively propagate through the interstellar medium, where they can be
either directly measured or indirectly
traced by means of \gr observations. Recent observations
made with the EGRET instrument on the Compton Gamma-Ray Observatory
of the diffuse Galactic \gr emission reveal a spectrum which is 
incompatible with the assumption that the cosmic ray spectra measured
locally hold throughout the Galaxy (Hunter et al. \cite{hu97}). The spectrum
observed with EGRET below 1 GeV is in accord with, and supports, the assumption
that the cosmic ray spectra and the electron-to-proton ratio
observed locally are uniform, however, the spectrum above 1 GeV, 
where the emission is supposedly dominated by $\pi^0$-decay, is harder
than that derived from the local cosmic ray proton spectrum.

In a recent paper Pohl \& Esposito (\cite{pe98}) demonstrated that if
the sources of cosmic rays are discrete, as are SNR, then
the spectra of cosmic ray electrons would vary and the locally measured 
electron spectrum would not be representative for the electron spectra
elsewhere in the Galaxy, which could be substantially harder than the local 
one. These authors have shown that the observed excess
of \gr emission above 1 GeV can in fact be explained as a correspondingly hard
inverse Compton component, provided the bulk of cosmic ray electrons is produced
in SNR.

In the following we will demonstrate that in case of a SNR origin
of cosmic ray nucleons part of the \gr excess may also be attributed to the 
dispersion of the spectral indices in these objects. In global averages, as 
are \gr line-of-sight integrals, this dispersion leads to a positive
curvature in the composite spectrum, and hence to modified $\pi^0$-decay
\gr spectra.

\section{The dispersion of cosmic ray spectra in SNR}
The synchrotron radio spectra from shell SNR indicate power law
spectra $I_{\nu }\propto \nu ^{-\alpha }$ (Clark \& Caswell \cite{cc76},
Milne \cite{m79}, Green \cite{gr00}) where the spectral index distribution 
from the sample of shell supernova remnants (SNR) has a mean value of 
about $<\alpha >\simeq 0.5$
and a significant standard deviation $\sigma _{\alpha }$. 
From synchrotron radiation
theory this implies a power law energy distribution of the radiating
relativistic electrons with mean spectral index $s_e=1+2<\alpha >\simeq 2.0$ and
dispersion $\sigma _e=2\sigma _{\alpha }$. Because the age of
the SNRs is much smaller than the characteristic radiative loss times of both cosmic ray nucleons and GeV electrons in
the remnant, and because both the particle acceleration processes
and the spatial propagation scale with rigidity, 
the relativistic hadrons should have the same momentum spectrum as
the relativistic electrons. After leaving their source these particles
propagate in the \ip by momentum-dependent spatial diffusion with a diffusion
coefficient $\kappa \propto p^b$, with $b\simeq 0.6$ as inferred from the
measurement of secondary to primary cosmic ray elements. Consequently, 
the source spectrum of cosmic ray hadrons 
\be 
N(p)=N_0({p\over mc})^{-s}
\label{hadronspec}
\ee
is a power law in momentum per nucleon with a spectral index having 
a mean value 
$<s>=1+b+2<\alpha >\simeq 2.7$
and dispersion $\sigma =\sigma _e$.

The Galactic cosmic ray hadron component results from the injection of 
relativistic
particles from many SNRs, especially in the inner part of the Galaxy.
In discussing the pros and cons of 
the Galactic origin of cosmic rays, Brecher and Burbidge (\cite{bb72}) noted
that the superposition of individual power law source spectra with dispersion
in the spectral index value displays a positive curvature in the total 
cosmic ray spectrum and in particular shows a flattening at higher energies, simply because those sources with the smallest spectral index 
dominate the total spectrum at large energies. If we represent the distribution
of hadron spectral indices by the Gaussian
\be
n(s)={1\over \sqrt{2\pi }\sigma }\; 
\exp \Bigl[-{(s-<s>)^2\over 2\sigma ^2}\Bigr]
\label{gauss}
\ee 
we obtain from Eq. (\ref{hadronspec}) for the averaged hadron spectrum 
\bdm
<N(p )>=\int_{-\infty }^\infty ds\ N(p ,s)\, n(s)
\edm
\be 
\hphantom{<N(p )>}
=N_0 \,\left({p\over {mc}}\right)^{-<s>+{{\sigma^2}\over 2} \ln\left({p\over {mc}}\right)}
\label{aver}
\ee 
By using $N(p)dp=N(\gamma )d\gamma $ and $p =mc\sqrt{\gamma ^2-1}$
we can determine from Eq. (\ref{aver}) the corresponding differential 
hadron spectrum with respect to the hadron Lorentz factor as
\be 
N(\gamma )=N_0 mc\; \gamma (\gamma ^2-1)^{{{-<s>-1}\over 2}+{{\sigma^2}\over 8} \ln\left(\gamma^2 -1\right)}\, 
\label{hadronlor}
\ee
With increasing hadron Lorentz factor $\gamma $ the energy spectrum 
(\ref{aver}) flattens, i.e. it hardens. Calculating the slope yields
\bdm 
\bar{s}\equiv -{\ln <N(\gamma )>\over \ln \gamma }
\edm
\be
\hphantom{\bar{s}}=
{<s>+1\over 2}{\ln (\gamma ^2-1)\over \ln \gamma }-\; 1
-\; {\sigma ^2 \over 8}{\ln ^2(\gamma ^2-1)\over \ln \gamma }
\label{slope}
\ee 
At relativistic energies Eq. (\ref{slope}) approaches 
\be
\bar{s}\simeq 
<s>-\; {\sigma ^2\over 2}\ln \gamma 
\label{sloperel}
\ee

\section{Pion decay gamma rays}
For the $\pi ^0 \to 2 \gamma $ decay the omnidirectional 
(i.e. integrated over the whole solid angle) differential $\gamma $-ray 
source function $q_{\pi ^0} (E_{\gamma }; \vec r)$ at the position 
$\vec r = (l,b,r)$ in space is related to the omnidirectional 
differential neutral pion source function
\be 
q_{ \pi ^0}(E_{ \gamma }; \vec r) = 2 \int_{U(E_{\gamma })}^{ \infty }
d\gamma _{ \pi } 
 {Q_{ \pi }(\gamma _{ \pi }; \vec r) \over [\gamma _{\pi }^2\ -\ 1]^{1/2}} 
\label{gammapi}
\ee 
with the lower integration boundary
\be
U(E_{\gamma })={E_{ \gamma }\over m_{ \pi }c^2}
 + {m_{ \pi }c^2\over 4E_{\gamma }}=f+\; {1\over 4f}
 \label{intbound}
 \ee
in terms of the 
dimensionless $\gamma$-ray photon energy
\be
f\equiv {E_{\gamma }\over m_{\pi }c^2}={E_{\gamma }\over 0.135 \; {\rm  GeV}}
\ee
in units of the pion rest energy.

The differential photon number flux of $\pi ^0$-decay $\gamma $-rays from a
 direction (l, b) is given by the line of sight integral of the 
 source function~(\ref{gammapi})
\be 
{dN_{\gamma } (E_{\gamma }; l, b) \over {dt \ dE_{\gamma } \ d \Omega }} 
= {1 \over 4 \pi } \int_0^{\infty } 
dr \ q_{\pi ^0} (E_{\gamma }; \vec r )
\label{gamflux}
\ee  
\subsection{Pion source spectrum}
The pion source spectrum has been frequently calculated on the basis of
the inclusive cross section and a delta-functional for the pion spectrum
(e.g Mannheim \& Schlickeiser \cite{ms94}). As we show in the Appendix 
this approximation may be useful at higher energies, but it is certainly 
inappropriate at energies not very far from the threshold.

Therefore in this section the pion source spectrum required in 
Eq.~(\ref{gammapi}) 
is calculated using cosmic hadron distribution function 
(\ref{aver}) as input for the Monte-Carlo code DTUNUC (V2.2) 
(M\"ohring \& Ranft \cite{MR91}, Ranft et al. 
\cite{RCT94}, Ferrari et al. \cite{Fer96a}, Engel et al. 
\cite{ERR97}),
which is based on a dual parton model (Capella et al. \cite{Cap94}).
This MC model for hadron-nucleus and nucleus-nucleus interactions
includes various modern aspects of high-energy physics and has been
successfully applied to the description of hadron production in
high-energy collisions (Ferrari et al. \cite{Fer96b}, Ranft \& Roesler
\cite{RR94}, M\"ohring et al. \cite{Moe93}, Roesler et al.
\cite{RER98}).

The \gr spectra thus derived can be compared with the observed
EGRET spectra from the inner Galaxy. Here we use the data of
all viewing periods of phases 1-4, corresponding to observations between
1991 April and 1995 October. From the observed intensity we subtract
the extragalactic background (Sreekumar et al. \cite{sre98})  and
the point-spread functions of all sources in the Third EGRET Catalogue
(Hartman et al. \cite{har98}). The intensity spectrum has additional
associated uncertainties, which arise from the subtraction of the 
extragalactic background and the sources. A systematical error of 10\% is
assumed to account for uncertainties in the spark chamber efficiency 
correction (Esposito et al. \cite{esp99}).

\begin{figure}
\resizebox{\hsize}{!}{\includegraphics{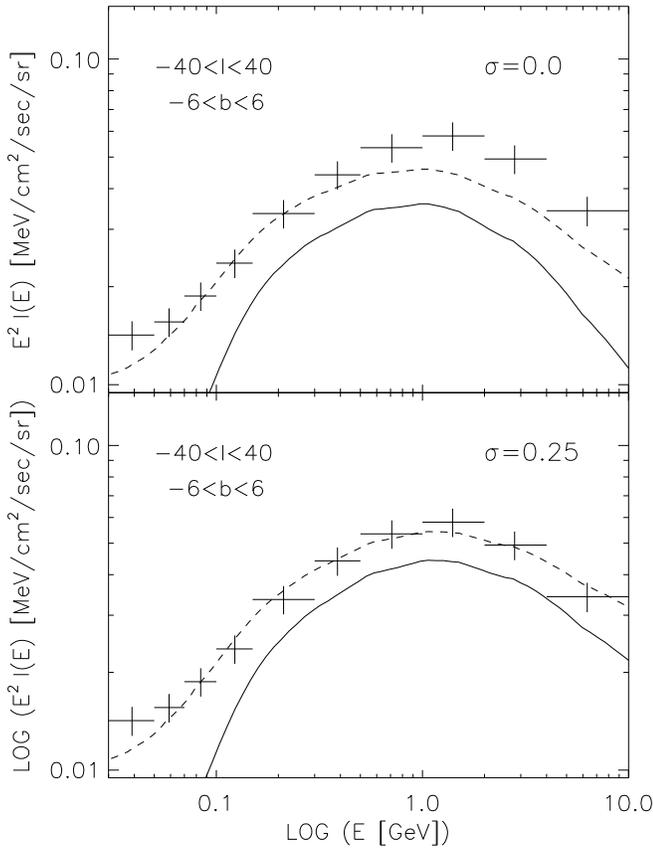}}
\caption{The observed intensity spectrum from the inner Galaxy shown
in comparison with $\pi^0$-decay spectra with (bottom panel) and without
(top panel) dispersion in the cosmic ray spectrum. The solid lines display the
$\pi^0$-decay spectra, and the dashed lines are the total \gr spectra including
the leptonic contribution, which here is simply given as a power law $\propto
E^{-2}$ with an intensity determined by the data for $E\le 100\,$MeV.
Whereas in the top panel, i.e. without dispersion, the GeV excess can be
clearly seen, the bottom panel proves that
a dispersion with $\sigma=0.25$ can in fact explain the data and
remove the GeV excess.}
\label{spektrum}
\end{figure}

The pion decay spectra are compared with the intensity of the diffuse
Galactic \gr emission from the inner Galaxy in Fig.~\ref{spektrum}. 
There the leptonic contribution to the diffuse \gr intensity has been
approximated by a simple power law $\propto E^{-2}$ with an intensity 
$E^2 I_l = 0.01 \ {\rm MeV/cm^2/sec/sr}$, adjusted to reproduce
the data for $E\le 100\,$MeV. It can be seen that 
a dispersion in the index of the cosmic ray source spectra of 
$\sigma\simeq 0.25$ is sufficient to explain the observed
intensity spectrum and remove the GeV excess. We can therefore conclude
that if dispersion is to be made responsible for the GeV excess,
it must be at a level of $\sigma\simeq 0.25$.

\section{Compatibility with upper limits at higher gamma ray energies}

A dispersion in the cosmic ray source spectra -- and thus in the cosmic ray
flux throughout the Galaxy -- has impact also on the resultant \gr
spectra at higher energies, both for the diffuse emission and for individual
supernova remnants. In this section we will investigate whether
a spectral dispersion of $\sigma=0.25$ is compatible with existing
upper limits for \gr emission in the TeV energy range. 

To date, observations of TeV \gr emission from individual SNR 
have yielded only a few detections, one of which (Cas~A, Aharonian et al.
\cite{aha01}) is presumably (Atoyan et al. \cite{ato00}) and the others 
are clearly caused by leptonic emission. Although deep surveys have
been performed, no 
hadronic TeV \gr emission has been unambiguously detected from SNR 
(Buckley et al. \cite{buc98}). These observational results have been a severe
constraint for simple and conventional models of cosmic ray acceleration in
SNR. If there were a dispersion in the cosmic ray spectra in SNR, some,
i.e. those with a hard spectrum, would be expected to be very prominent 
TeV sources. Therefore the observational constraints are even more severe
if there is a noteworthy dispersion in the particle spectra. 

However,
these constraints rely on the assumption that the power law spectra of cosmic
rays in the SNR persist to very high energies like $10^{14}-10^{15}\ $eV.
This may not be a valid assumption; in fact more detailed models of
particle acceleration including non-linear effects indicate that
the resultant particle spectra may show cut-offs at energies much smaller
than $10^{14}\ $eV (Baring et al. \cite{bar99}). There is also observational 
evidence that SNR do not produce straight power laws up to $10^{15}\ $eV.
For all SNR the X-ray flux, synchrotron or not, 
is less than the extrapolated radio synchrotron spectrum. Since many
of the sources, in particular the five historical remnants, are too young for
the electron spectra to be limited by energy losses, acceleration cut-offs 
must occur at electron energies of 100 TeV or less (Reynolds \&
Keohane \cite{rk99}). At these energies the acceleration process should 
operate similarly for electrons and ions, and similar cut-offs must be
expected for the cosmic ray nuclei, which would be intrinsic to the actual acceleration process. Once a remnant is in the Sedov phase, the maximum energy
of accelerated particles rises very little. Cas A is probably still in the free
expansion phase, but proper-motion measurements of Kepler, Tycho and SN1006
indicate that these objects are well in the Sedov dynamics
(Moffett, Goss and Reynolds \cite{mgr93}). These remnants are therefore
unlikely to accelerate cosmic rays to energies higher than $10^{14}\ $eV at
any time.

We have made a statistical
argument in this paper and therefore may not discuss individual sources,
but generally speaking it would seem that a dispersion in the spectra
of cosmic rays in SNR is not in conflict with the existing upper limits for 
TeV \gr emission, when an upper cut-off in the cosmic ray spectra is taken into
account. The same argument applies to the Galactic diffuse \gr emission
at TeV energies as we discuss in somewhat more detail.
  
The Whipple collaboration has recently published an upper limit for
\gr emission at 500 GeV from the Galactic plane, based on observations
in a small area in the plane defined by $38.5\degr \le l\le 41.5\degr$
and $-2\degr \le b\le 2\degr$  (LeBohec et al.~\cite{leboh00}). 
An upper limit for
\gr emission at 1\,TeV from the Galactic plane ($38\degr \le l\le 43\degr$,
 $-2\degr \le b\le 2\degr$) 
has been reported by the HEGRA  collaboration (Aharonian et al.  \cite{aha01a}).
Together with
the \gr intensity in the GeV range these observations constrain models of
the GeV excess. Composite spectra of diffuse
Galactic \gr emission from 1 GeV to 500 GeV can be compiled, however, they are
subject to systematic uncertainties for a number of reasons. 
The numerical value of this upper limit at 500 GeV depends to some extent on 
the \gr spectrum itself.
Also the observed portion of sky is so small that the finite extent of the
EGRET point-spread function becomes non-negligible. To derive such a composite
spectrum of diffuse \gr emission we will therefore use EGRET data for 
energies above 1 GeV only, for which the 68\% containment radius is 
$\lesssim 1.7\degr$. To account for the effect of the point-spread function 
and to increase the number of counts, we will also derive the EGRET 
intensity spectrum from
an area slightly larger than that observed with Whipple, that is
$37\degr \le l\le 43\degr$ and $-2.5\degr \le b\le 2.5\degr$. When comparing the
$\pi^0$-decay spectrum with the composite spectrum of diffuse Galactic \gr 
emission, it may be sufficient to use an analytical approximation for the
differential pion production cross section, because the uncertainty thus 
imposed is probably not larger than the systematic uncertainty in the data,
but the procedure is substantially simplified. Our analytical approximation 
for the calculation of the pion source spectra is described in Appendix A.

Given the \gr intensity at GeV energies, the model intensity at 500 GeV
depends on a) the mean proton spectral index (fixed to $<s>=2.7$
in this study), b) a possible dispersion in the spectral indices of the 
cosmic ray sources ($\sigma = 0.25$ is required to account for the GeV excess),
and c) the limiting energy to which the power law spectra of the cosmic ray 
sources persist. Since we want to investigate whether or not the
required dispersion $\sigma = 0.25$ is compatible with the Whipple upper limit,
the only number which we may vary is the high energy cut-off in the 
proton spectrum. Therefore the question is: what is the high energy cut-off
$\gamma_{\rm max}$ required to satisfy the Whipple upper limit at 500 GeV,
given the observed GeV intensity, $<s>=2.7$, and $\sigma = 0.25$. Then
we may ask whether or not $\gamma_{\rm max}$ thus derived is reasonable
in view of the cosmic ray all-particle spectrum and the observed composition
near and at the knee.

\begin{figure}
\resizebox{\hsize}{!}{\includegraphics{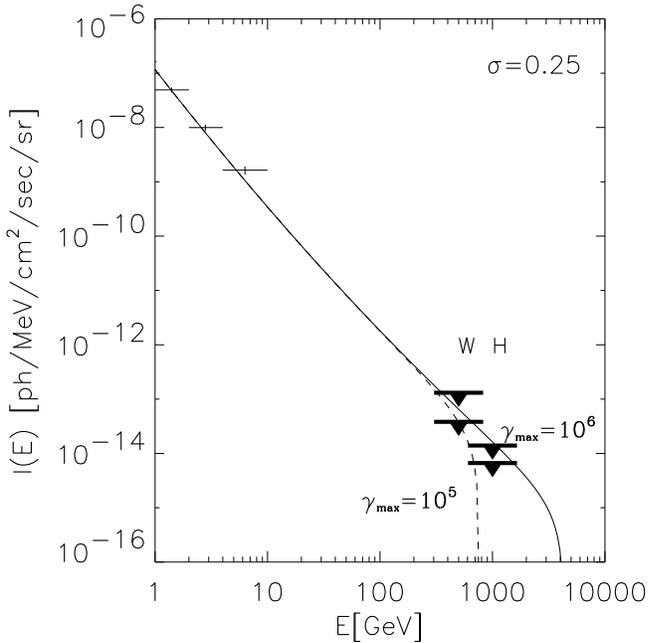}}
\caption{Composite spectrum between 1 GeV and 1 TeV
of the diffuse Galactic \gr emission from the Galactic plane at 
$l\approx 40\degr$ based on EGRET, Whipple and HEGRA (Aharonian et al. \cite{aha01a}) data. 
The upper limit at 500 GeV (Whipple), which is
at the 99.9\% confidence level, depends on the \gr spectrum. The lower mark
applies for a \gr spectral index $s=2.0$ whereas the upper mark is appropriate
for a spectral index $s=2.6$. 
The limit at 1\,TeV (HEGRA), which is a the 99\%   confidence level, 
is plotted for two different 
ways of background subtraction (see  Aharonian et al. \cite{aha01a}).
The data are compared with $\pi^0$-decay spectra 
calculated for a cosmic ray spectrum produced by individual sources with
mean spectral index $<s>=2.7$ and dispersion $\sigma =0.25$. The solid line 
refers to a high energy cut-off in the cosmic ray proton spectrum at
$\gamma_{\rm max}=10^6$. It should be compared with the lower mark  for the
upper limit at 500 GeV which it clearly and significantly exceeds. It also 
exceeds the HEGRA upper limits at 1\,TeV for both ways of background subtraction.
For $\gamma_{\rm max}=10^5$ the dashed line shows no contradiction with the 
Whipple and HEGRA data, when compared to the upper 
mark for the upper limit at 500\,GeV.
}    
\label{upplimtest}
\end{figure}

As can be seen in Fig.~\ref{upplimtest}, a dispersion of $\sigma=0.25$ 
clearly violates the upper limit at 500 GeV derived by the Whipple team, if
$\gamma_{\rm max}=10^6$, but the upper limit is clearly satisfied, if
$\gamma_{\rm max}=10^5$. Therefore, if a spectral dispersion exists at the
level required to explain the GeV excess, there must be a high energy cut-off 
in the cosmic ray source spectra somewhere between $\gamma_{\rm max}=10^5$
and $\gamma_{\rm max}=10^6$.

\section{Discussion}
In this paper we have calculated the high energy diffuse Galactic
\gr emission produced by the hadronic component of cosmic rays
under the assumption that the interstellar cosmic ray spectrum is a superposition of individual power-law spectra with a dispersion in the 
spectral index as a result of cosmic ray production in SNR.
We have shown that a diffuse \gr spectrum thus derived can in fact
explain the GeV excess, provided the dispersion in the individual SNR
production spectral indices is $\sigma =0.25$.

We have then investigated under what assumptions a composite cosmic ray
spectrum with $\sigma =0.25$ would be compatible with data in the TeV 
energy range both for individual SNR and for the diffuse emission. It is
found that the high energy data would require the existence of a 
high energy cut-off in the cosmic ray source spectra at proton energies not higher than somewhere between Lorentz factors $\gamma_{\rm max}=10^5$
and $\gamma_{\rm max}=10^6$.

The hypothesis presented here provides a possible explanation for the GeV 
excess, as do other models. 
It has a number of consequences for observable quantities
which can be used to test the viability of the scenario and
which we want to discuss in this section.

Is a dispersion of $\sigma =0.25$ compatible with the radio synchrotron
spectra of SNR? The following discussion applies only to shell-type SNR
which in fact have an average synchrotron power index $\alpha\simeq 0.5$
(Green \cite{gr00}).
This translates to an electron spectral index $s\simeq 2$.
For various reasons the uncertainties in the derived radio spectral indices
of SNR are often large, but nevertheless some shell-type SNR show
spectral indices which are significantly different from 0.5, either flatter
or steeper than the average spectrum. The actual distribution of observed
spectral indices results from both an intrinsic dispersion and the
observational uncertainties, and it is not a trivial task to deduce
the intrinsic dispersion. We have taken the radio spectral index data for 
Galactic shell SNR given in Fig.5 of Green (\cite{gr00}) and have performed 
a $\chi^2$-test to see whether the sample of measurements is consistent with
a uniform source spectral index. 
The answer is clearly "no", the reduced $\chi^2$ is 
around 12. We have then approximated the effect of a distribution of source
spectral indices by adding a dispersion $\sigma_\alpha$ to the 
measurement errors.
A reduced $\chi^2$ of unity results when the dispersion is 
$\sigma_\alpha=0.10$, then
also $<\alpha>=0.53\pm 0.02$. Taking out Cas A, which has a soft spectrum 
with very small error bars, gives 
$\sigma_\alpha=0.085$ and $<\alpha>=0.52\pm 0.02$.
This analysis indicates that a dispersion in the radio spectral indices of
shell SNR exists, but only at a level of $\sigma_\alpha =0.085-0.1$, corresponding to $\sigma_s =0.17-0.2$ in the cosmic ray electron spectra.
This falls only marginally short of the $\sigma=0.25$ which is 
required to explain the GeV excess. Therefore, given the uncertainties, we
conclude that $\sigma =0.25$ is compatible with the radio synchrotron
spectra of SNR.

Why don't we observe a dispersion in the local cosmic ray spectrum? 
Here it is useful to calculate the number of SNR which would
actually contribute to the local cosmic ray flux. The local supernova rate
is about $S=30$ Myr$^{-1}\,$kpc$^{-2}$. In standard cosmic ray diffusion models
(e.g. Webber, Lee and Gupta \cite{wlg92}) the range of the cosmic rays is linked
to the halo size $H\simeq 3\ $kpc and the life time is identical to the escape
time $\tau \simeq 20\,\gamma^{-0.6}\ $Myr. The number of SNR contributing to the
local cosmic ray (proton) flux is then $N=\pi\,H^2\,\tau\,S\simeq
17000\,\gamma^{-0.6}$.
In other words, while at a proton energy of 50 GeV some 1600 SNR would 
contribute, at 100 TeV the number would be down to 17 SNR.
 Even if this is a rather rough estimate (using an averaged diffusion 
coefficient for halo and Galactic disc and neglecting reacceleration), it shows clearly that the
number of SNR contributing to the local CR flux becomes  small at energies
of some 100\,TeV in which case the local
cosmic ray spectrum could strongly deviate from the average spectrum in 
Eq.~\ref{hadronlor}. A similar effect affects the cosmic ray electron spectrum
at much smaller energies of 100 GeV, which is the basis for the inverse Compton
models of the GeV excess (Pohl \& Esposito \cite{pe98}). 

Added to this would be the effect of deviations from power law behaviour in
the cosmic ray source. We have found that there must be a cut-off at proton 
energies not higher than a few hundred TeV. This can be established as
sharp cut-offs, possibly with a distribution of cut-off energies, or also 
in the form of a spectral steepening at somewhat smaller energies,
which behaviour is predicted in models of non-linear shock acceleration
(e.g. Baring et al. \cite{bar99}). As a result 
the average spectrum at higher energies would be softer than 
what is given in Eq.~\ref{hadronlor}, which would not change the expected
GeV \gr spectrum, but the local cosmic ray proton spectrum above 10 TeV,
which is observed to be slightly softer than that obtained by direct measurements at lower energies (Asakimori et al. \cite{asa98};
Amenomori et al. \cite{ame01}). All in all the lack of dispersion in the
local cosmic ray proton spectrum does not seem to contradict the hypothesis
presented in this paper.

Is the high energy limit in the cosmic ray proton source spectra compatible
with the observed all-particle spectrum and the composition near the knee?
We have seen that the local cosmic ray spectrum near the knee would be produced
by few nearby SNR which would cause a structured
spectrum near the knee (Erlykin \& Wolfendale \cite{ew97a,ew97b,ew98a,ew98b}).
We have not investigated the composition and in particular not studied the 
possibly different source spectra for different species. Therefore we feel
not well equipped to discuss the issue of cosmic ray composition near the knee.
The fact that a softening in the cosmic ray source spectra is required may
indicate a potential problem with the notion that SNR accelerate cosmic rays 
up to the knee, but a more detailed and careful study will be needed to address
this issue.

Since there is observational evidence for a dispersion in the spectral indices
of the cosmic ray spectra in SNR, the composite cosmic ray spectrum must be 
curved, if the cosmic rays are predominantly produced in SNR. We have shown that
if the dispersion is as strong as $\sigma=0.25$, its effect on the interstellar
cosmic ray spectrum would explain the GeV excess in the diffuse Galactic
\gr spectrum. If the actual dispersion is weaker than this, it would still
contribute to the GeV excess and therefore should not be neglected. There
are other viable models for the GeV excess and it may well be
that it is a combined effect of the spectral dispersion and, e.g., a hard 
inverse Compton spectrum (Pohl \& Esposito \cite{pe98}).

\appendix

\section{Approximative calculation of pion spectra}
\label{appa}
We consider production of $\pi^0$s by cosmic rays via the 
reaction p+p$\to \pi ^0 + X$, where $X$ stands for any other particle,
as described in Mannheim \& Schlickeiser (\cite{ms94}).
The total cross section for pion production is  
$\sigma_{\mathrm{pp}}^{\pi}\simeq \sigma_{\mathrm{p,inel}}\simeq 3\cdot 10^{-26}$\,cm$^2$.
For charged pions the mean multiplicity has an energy
dependence $\xi \simeq 2 E_p^{1/4} \ ({\rm GeV})$ where $E_p$ is in GeV. 
Due to isospin symmetry $\xi_{\pi^0}={1\over 2}\,\xi_{\pi^\pm}$.
Laboratory measurements of the mean pion energy in p-p and p-$\alpha $ 
interactions indicate that a constant fraction, about 30$\% $, of the incident 
kinetic energy of protons goes to pion energy. This is in agreement 
with Fermi's theory of pion production, in which a thermal equilibrium 
is assumed in the resulting pion cloud and in which the mean Lorentz 
factor of pions in the laboratory frame (where one of the protons is 
initially at rest) is given by
\be
\bar {\gamma }_{\pi }\simeq \gamma _p^{3/4} 
\label{mean}
\ee
where $\gamma _p$ denotes the Lorentz factor of the incident proton. 

We approximate the differential cross-section by 
the experimentally measured inclusive cross-section multiplied with 
a $\delta $-function centered at the mean pion energy (\ref{mean}), that is
\begin{eqnarray}
\sum_k \sigma _{k, pp, p \alpha }^{ \pi ^0 } (\gamma _{\pi }, \gamma _p) 
& \simeq &  \xi \sigma _{pp, p \alpha }^{\pi ^0} (\gamma _p) \ 
\delta (\gamma _{\pi }- \bar {\gamma }_{\pi }) 
\nonumber
\\
&\simeq& \xi \sigma _{pp, p \alpha }^{\pi ^0} (\gamma _p)  \ 
\delta ( \gamma _{\pi } -  \gamma _p^{3/4}) 
\label{crossapp}
\end{eqnarray}
This $\delta $-function approximation and the use of the 
same mean pion energy (\ref{mean}) in p-p and p-$\alpha $ collisions do not 
introduce appreciable error, provided the incident proton is highly 
relativistic so that the scaling relations apply. At energies much higher than
$10^4$ GeV Eq.~(\ref{crossapp}) overestimates the number of pions, but at these
energies \grs are also efficiently produced by kaons, so that the use of
Eq.~(\ref{crossapp}) will not lead to a significant overestimate of the 
\gr source function.
(Dermers (\cite{d86})  formula yields an inclusive cross section 
20\% smaller than that given by  Eq.~(\ref{crossapp}), with an average pion 
energy roughly 50\% higher than given by
 Eq.~(\ref{crossapp}), the 
energy 
distribution 
of secondary pions going up almost to the kinetic energy of the incident 
proton,  
overpredicting pions at higher rapidities (Mori \cite{m97}).
Using Dermers formula, the cut off in the \gr spectrum becomes smoother
so the required cut off energy in the CR proton spectrum is  smaller than
in our conservative calculation.)

For the pion power we then obtain 
\begin{eqnarray}
P (\gamma _{\pi }, \gamma ) & =& 1.3 \ \beta c \gamma _{\pi } 
\ [n_{HI} ( \vec r) + 2n_{H_2} ( \vec r ) ] 
\,\xi \sigma _{pp}^{\pi } (\gamma ) 
\label{powerpi2}\nonumber
\\
& & \times \, \delta ( \gamma _{\pi } - \gamma ^{3/4})\, H [\gamma  - \gamma _{th}] 
\end{eqnarray}
where the factor 1.30 accounts for the known chemical composition of the 
interstellar medium.  $\gamma _{th}$ is the hadron
threshold Lorentz factor. 
The pion source function then is
\begin{eqnarray}
Q_{\pi } (\gamma _{\pi }; \vec r ) & = & {1.26 \over \gamma _{\pi }m_{\pi }c^2} 
\int_1^{\infty } d \gamma\  N ( \gamma , \vec r ) 
\,P_{\pi ^0} (\gamma _{\pi }, \gamma ) \nonumber \\
& = & {1.64  \over m_{\pi }c}\, [n_{HI} ( \vec r ) + \ 2n_{H2} ( \vec r )] 
\ \times\nonumber \\
& & \int_{\gamma _{th}}^{\infty } d \gamma\ N ( \gamma , \vec r )\,
\beta\, \xi \sigma _{pp}^{\pi ^0} (\gamma )\, \delta ( \gamma _{\pi } - \gamma ^{3\over 4}) 
\end{eqnarray} 
where the factor 1.26 accounts for the contribution of $\alpha $-p, 
$\alpha - \alpha $ collisions and collisions of higher metallicity 
cosmic rays, and is derived using the ratio of the respective 
inclusive cross-section to the p-p inclusive cross-section and the 
known elemental composition of the cosmic rays.

For a separable differential number density distribution 
of cosmic ray hadrons $N( \gamma , \vec r )= N_0(\vec r)\,N(\gamma )$ 
at position $ \vec r = (l, b, r)$ in the Galaxy and a spectrum as given in
Eq.~(\ref{hadronlor}) with a high energy cut-off at $\gamma _{\rm max}$ 
we obtain for relativistic particles ($\beta\simeq 1$)
\begin{eqnarray}
Q_{\pi } (\gamma _{\pi }; \vec r )&=&
{2.2 \ m_{\rm p} \over m_{\pi } } 
\ N_0 ( \vec r )\  [n_{HI} ( \vec r ) + 2n_{H_2} ( \vec r )]\ 
\gamma _{\pi }^{1/3}\ \times
\nonumber \\
&&\xi \sigma _{pp}^{\pi^0}(\gamma _{\pi }^{4/3})\  (\gamma _{\pi }^{8/3}-1)^{-{{<s>}\over 2}+
{{\sigma^2}\over 8}\, \ln (\gamma _{\pi }^{8/3}-1)}\  
\nonumber \\
&& \qquad
{\rm for}\quad 1\ll \gamma_\pi \le \gamma_{\rm max}^{3/4}
\label{pionsource}
\end{eqnarray}
so that with the assumptions made above, the pion source 
function~(\ref{pionsource}) reduces to
\begin{eqnarray}
 Q_{\pi } (\gamma _{\pi }; \vec r ) &\simeq& 
  {6.6\cdot 10^{-26} \ m_p \over m_{\pi }} 
\ N_0 ( \vec r )\, [n_{HI} ( \vec r ) + 2n_{H_2} ( \vec r )] 
\, \nonumber \\ &&
 \times \, \gamma _{\pi }^{{{2-4<s>}\over 3}+{{8\sigma^2}\over 9}\, 
\ln(\gamma_\pi)}   \nonumber \\ &&
 \qquad
{\rm for}\quad 1\ll \gamma_\pi \le \gamma_{\rm max}^{3/4}
\label{piso}
\end{eqnarray}
Eq.~(\ref{piso}) can then be used to calculate the \gr source function
in Eq.~(\ref{gammapi}).

\begin{acknowledgements}
Partial support by the Bundesministerium f\"ur Bildung und Forschung through 
the DLR, grant {\it 50 OR 0006}, is gratefully acknowledged.
\end{acknowledgements}

{}

\end{document}